\documentclass[aip,preprint]{revtex4-1}
\usepackage{graphicx}
\usepackage{bm}
\usepackage{color}

\begin{document}

\preprint{AIP/123-QED}

\title{Electron-magnetohydrodynamic simulations of electron scale current sheet dynamics in the Vineta.II guide field reconnection experiment}
\affiliation{Max Planck Institute for Solar System Research,
Justus-von-Liebig-Weg 3, 37077, G\"ottingen, Germany.}
\affiliation{Max Planck Institute for Plasma Physics, Wendelsteinstr. 1, 17491 Greifswald, Germany.}
\author{Neeraj Jain}
\affiliation{Max Planck Institute for Solar System Research,
Justus-von-Liebig-Weg 3, 37077, G\"ottingen, Germany.}
\affiliation{Max Planck/Princeton Center for Plasma Physics.}
\author{Adrian von Stechow}
\affiliation{Max Planck Institute for Plasma Physics, Wendelsteinstr. 1, 17491 Greifswald, Germany.}
\affiliation{Max Planck/Princeton Center for Plasma Physics.}
\author{Patricio A. Mu\~noz}
\affiliation{Max Planck Institute for Solar System Research,
Justus-von-Liebig-Weg 3, 37077, G\"ottingen, Germany.}
\affiliation{Max Planck/Princeton Center for Plasma Physics.}
\author{J\"org B\"uchner}
\affiliation{Max Planck Institute for Solar System Research,
Justus-von-Liebig-Weg 3, 37077, G\"ottingen, Germany.}
\affiliation{Max Planck/Princeton Center for Plasma Physics.}
\author{Olaf Grulke}
\affiliation{Max Planck Institute for Plasma Physics, Wendelsteinstr. 1, 17491 Greifswald, Germany.}
\affiliation{Max Planck/Princeton Center for Plasma Physics.}
\author{Thomas Klinger}
\affiliation{Max Planck Institute for Plasma Physics, Wendelsteinstr. 1, 17491 Greifswald, Germany.}

\date{\today}

\begin{abstract}
Three dimensional electron-magnetohydrodynamic (EMHD) simulations of electron current sheet dynamics in a background of stationary and unmagnetized ions and the subsequent generation of electromagnetic fluctuations are carried out.
The physical parameters and initial magnetic configuration in the simulations are chosen to be similar to those in the \textsc{Vineta}.II magnetic reconnection experiment. 
Consistent with the experimental results, our 3D EMHD simulations show the formation of an elongated electron scale current sheet together with the excitation of electromagnetic fluctuations within this sheet. The fluctuations in the simulations are generated by an electron shear flow instability growing on the in-plane (perpendicular to 
the direction of the main current in the sheet) electron shear flow (or current) developed during the current sheet evolution. Similar to the experiments, the magnetic field fluctuations perpendicular to the guide magnetic field 
exhibit a broadband frequency spectrum following a power law and a   positive correlation with the axial current density. 
Although the experimental results show that ions influence the spectral properties of the fluctuations, the simulations suggest that  the electron dynamics, even in the absence of ion motion, primarily determines the formation of the current sheet and the generation of electromagnetic fluctuations observed in the experiments.
\end{abstract}

\pacs{Valid PACS appear here}

\maketitle

\section{\label{sec:introduction}Introduction}
Magnetic reconnection releases the energy stored in magnetic fields in the form of kinetic energy and heat in a wide variety of plasma phenomena including solar flares, substorms in Earth's magnetosphere, sawtooth crashes in tokamaks and many astrophysical systems, e.g., accretion disks. This requires topological changes of magnetic field lines, enabled by dissipation in highly localized current sheets which form in these plasmas \cite{yamada2010}. In collisionless systems, the dissipation region where magnetic field lines break and reconnect becomes very thin and typically develops a two-scale structure: An electron current sheet with a thickness of the order of an electron inertial length $d_e=c/\omega_{pe}$ (for weak guide field) or electron thermal gyro-radius $\rho_{e}=v_{the}/\omega_{ce}$ (for large guide field) embedded within an ion current sheet with a thickness of the order of an ion inertial length $d_i=c/\omega_{pi}$ (for weak guide field) or ion thermal gyro-radius $\rho_{i}=v_{thi}/\omega_{ci}$ (for large guide field). Here the guide field is an external magnetic field in the direction of the current, and  $c$, $v_{the,i}$, $\omega_{pe,i}$ and $\omega_{ce,i}$  are the speed of light, thermal speed, plasma frequency and cyclotron frequency respectively. The subscripts 'e' and 'i' in these symbols represent electrons and ions, respectively. An effective dissipation that leads to the reconnection of field lines is provided by microphysical plasma processes at these respective scales.

These current sheets are susceptible to a variety of instabilities which can generate microturbulence in the dissipation region. Laboratory experiments \cite{ji2004,fox2010,inomoto2013,dorfman2014,kuwahata2014,stechow2016} and {\it in-situ} space observations \cite{eastwood2009,zhou2009,retino2007} of magnetic reconnection often show electromagnetic and electrostatic fluctuations as signatures of these instabilities: Large amplitude whistler waves along with ion acoustic and Langmuir turbulence were observed in an early reconnection laboratory experiment by Gekelman and Stenzel \cite{gekelman1984}. In the Magnetic Reconnection Experiment (MRX), electrostatic \cite{carter2002} and electromagnetic \cite{ji2004} fluctuations in the lower hybrid frequency $\omega_{lh}\approx \sqrt{\omega_{ci}\omega_{ce}}$ range were observed at the edge (low plasma $\beta$) and center (high plasma $\beta$) of the current sheet, respectively, where 
plasma $\beta$ is defined as the ratio of the plasma and magnetic pressure. Electrostatic fluctuations in the same frequency range were also observed in driven reconnection experiments at the Versatile Toroidal Facility (VTF), along with high frequency Trivelpiece-Gould wave turbulence \cite{fox2010}. Similar observations were made by the Cluster spacecraft in Earth's magnetotail: electrostatic and electromagnetic fluctuations near the lower hybrid frequency were observed during the crossing of the separatrix and the center of the current sheet, respectively \cite{zhou2009}. Large amplitude electromagnetic fluctuations in the ion cyclotron frequency range with properties similar to kinetic Alfv\'en waves (KAW) propagating obliquely to the guide field  were observed near the X-point in the TS-3 device \cite{inomoto2013}. In the \textsc{Vineta}.II guide field magnetic reconnection experiment, broadband whistler-like electromagnetic fluctuations somewhat centered around the lower hybrid frequency were observed near the X-point  in an electron scale current sheet \cite{stechow2016}.

The nature of the fluctuations and associated instabilities depend on the plasma parameters and magnetic field configurations which vary widely among laboratory experiments and space observations. In this paper, we focus on the plasma and field configurations of the \textsc{Vineta}.II magnetic reconnection experiment \cite{stechow2016}. In these experiments, electromagnetic fluctuations develop in a self-consistently formed electron current sheet. The spatial (half thickness of the electron current sheet $\sim 5\,d_e$) and temporal scales (angular frequency $ \omega \sim \omega_{lh} > \omega_{ci}$)
of the dynamics of the electron current sheet in the experiments are between ion and electron scales. Although electrons are expected to be the dominant contributors to the dynamics at these scales, ions can also influence the dynamics. Experimental results also show a dependence of the fluctuations on the ion mass. In the present study, however, we neglect ion dynamics and use an electron-magnetohydrodynamic (EMHD) model as a first step towards understanding the underlying physical processes. In fact, the purpose here is not to reproduce the exact experimental results but to identify the minimal physics of the current sheet dynamics and subsequent generation of the electromagnetic fluctuations in the \textsc{Vineta}.II experiments.




The remainder of this paper is organized as follows: Section \ref{sec:emhd} introduces the EMHD model and its associated current sheet instabilities. 
This is followed by a short summary of key experimental results observed in the \textsc{Vineta.II} experiment in section \ref{sec:keyresults}. The detailed simulation setup is described in section \ref{sec:simu_setup}, followed by its results and their comparison to experimental observations in section \ref{sec:results}, specifically addressing current sheet formation and fluctuation dynamics. Finally, section \ref{sec:conclusion} interprets the simulation results in terms of their applicability to the experimental situation and identification of the underlying instability.

\section{Electron-magnetohydrodynamic (EMHD) model and electron shear flow instabilities\label{sec:emhd}}

\subsection{EMHD model}
The EMHD model is a fluid description for electron dynamics in a stationary background of ions. It is valid for spatial scales smaller than $d_i$ and time scales smaller than $\omega_{ci}^{-1}$. The electron dynamics is described by the electron momentum equation coupled to Maxwell's equations. An equation for the evolution of the magnetic field $\mathbf{B}$ can be obtained by eliminating the electric field from the electron momentum equation by using Faraday's law \citep{kingsep90},
\begin{eqnarray}
\frac{\partial}{\partial t}(\mathbf{B}-d_e^2\nabla^2\mathbf{B})&=&\nabla \times
[\mathbf{v}_e\times (\mathbf{B}-d_e^2\nabla^2\mathbf{B})]
\label{eq:emhd1},
\end{eqnarray}
where $\mathbf{v}_e=-(\nabla\times\mathbf{B})/\mu_0n_0e$ ($\mu_0$, $n_0$ and $e$ are the permeability of the vacuum, electron density and elementary charge, respectively) is the electron fluid velocity and $d_e=c/\omega_{pe}$ is the electron inertial length. 
With stationary ions, the current density is given by $\mathbf{j}=-n_0e\mathbf{v}_{e}$. In addition to the neglect of  ion dynamics, Eq. (\ref{eq:emhd1}) assumes uniform electron number density $n_0$ and incompressibility of the electron fluid, while the displacement current is ignored under the low frequency assumption $\omega \ll \omega_{pe}^2/\omega_{ce}$. In the absence of electron inertia ($d_e\rightarrow 0$), Eq. (\ref{eq:emhd1}) implies frozen-in flux, i.e. the magnetic field is rigidly coupled to the electron fluid. Bulk electron inertia ($m_e$), contained in the definition of $d_e \propto \sqrt{m_e}$, provides a non-ideal effect in the generalized Ohm's law that breaks the frozen-in condition of magnetic flux. Its inclusion extends the validity of the EMHD model to spatial scales smaller than $d_e$. 

In this publication, we aim to identify a minimal model of the current sheet dynamics in the experiments. 
We, therefore, keep the velocity and magnetic field gradients in our model but ignore the density gradients  and finite resistivity which are present in the experiments. In EMHD, gradients in the electron flow velocity, but not in density, provide a free energy source for the current sheet instabilities to grow \cite{sundar2010}. While density gradients can in principle trigger velocity gradient driven instabilities even in the absence of electron inertia \cite{wood2014}, we  focus instead on flow gradient driven instabilities triggered by electron inertia in the collisionless limit.

\subsection{Electron shear flow instabilities}
As the name suggests, electron shear flow instabilities (ESFI) feed on gradients in electron flow velocity and are triggered by electron inertia. In collisionless magnetic reconnection, 
an electron shear flow develops where the ion flow velocity is negligible compared to that of the electrons. This flow amounts to an electron current sheet with a thickness (or shear length scale) $\delta$ of the order of several $d_e$.
Such a current sheet forms in \textsc{Vineta.II} experiments of magnetic reconnection. 
Depending on $\delta$ and the external guide magnetic field strength in the electron current direction, the spontaneous growth of the ESFI can be dominated either by tearing or non-tearing modes: The former dominates for small guide fields (smaller than the edge magnetic field $B_{edge}$ of the current sheet) and a current sheet half thickness close to $d_e$ \cite{jain2015}.

In the \textsc{Vineta.II} reconnection experiments, the half thickness ($\sim 5\,d_e$) of the electron current sheet and the strength of the guide field ($\sim 10\, B_{edge}$) favor the growth of the non-tearing mode. For these parameters, the growth rate $\gamma_f$ and the wave numbers, parallel ($k_z$) and perpendicular ($k_y$) to the guide magnetic field,  of the fastest growing non-tearing mode are  $\gamma_f \sim 0.01\,\omega_{ce}$ ($\omega_{ce}=eB_{edge}/m_e$), $k_yd_e\sim 10$ and $k_zd_e\sim 1$ \cite{jain2015}. Unlike the tearing mode, the linear eigenfunction $\tilde{v}_x/v_{Ae}$ (where $v_{Ae}=d_e\omega_{ce}$ is the electron Alfv\'en velocity) of the non-tearing mode does not change sign across the shear layer and is typically significantly larger than the eigenfunction $\tilde{b}_x/B_{edge}$. Here, $\tilde{b}_x$ and $\tilde{v}_x$ are the components of the perturbed magnetic field and electron flow velocity in the direction of the shear. 
The nonlinear evolution of the non-tearing modes shows the development of characteristic electron flow vortices within the shear layer \cite{jain2017}.

\section{Key experimental results}
\label{sec:keyresults}

\begin{figure}[h]
\includegraphics[width=0.65\textwidth]{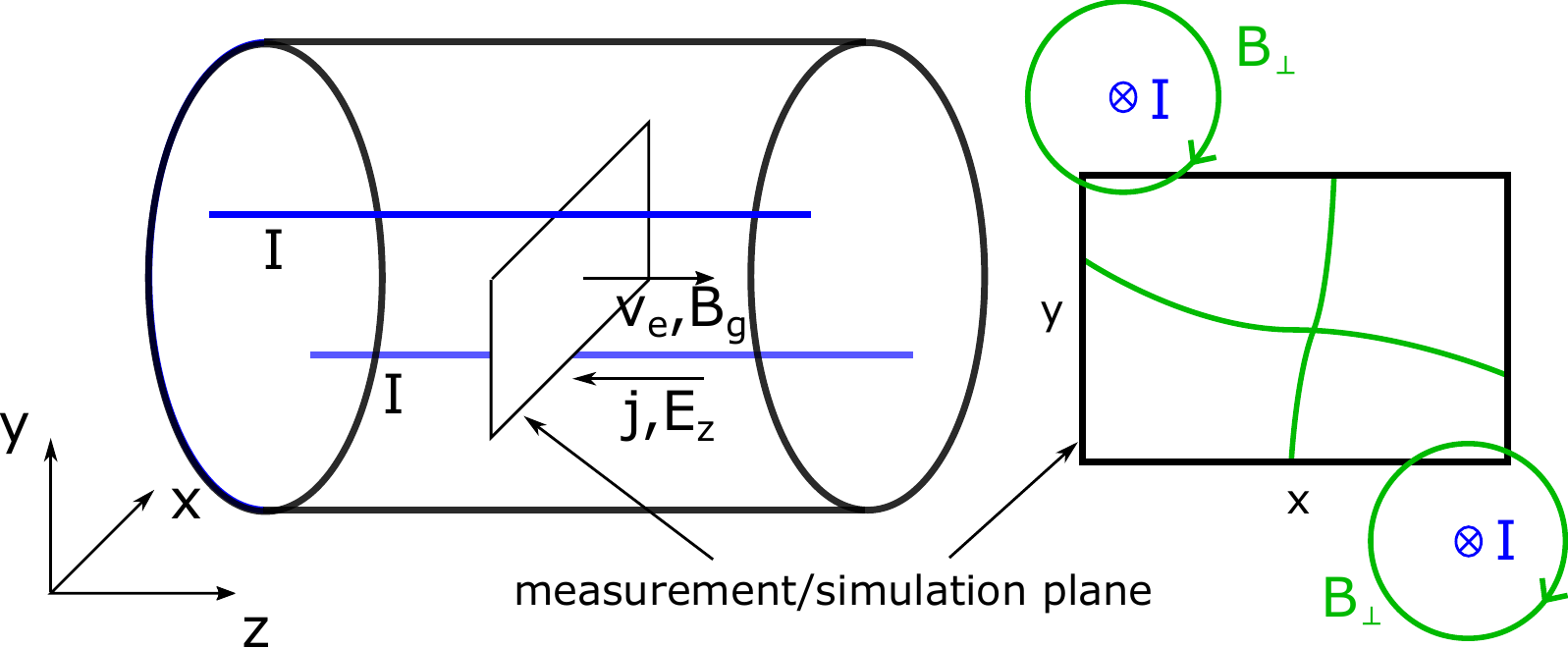}
\caption{Experiment overview (left) and measurement/simulation plane (right).}
\label{fig:exp-overview}
\end{figure}

\begin{figure}[h]
\includegraphics[clip,width=0.48\textwidth,trim=1cm 0.3cm 1cm 1cm]{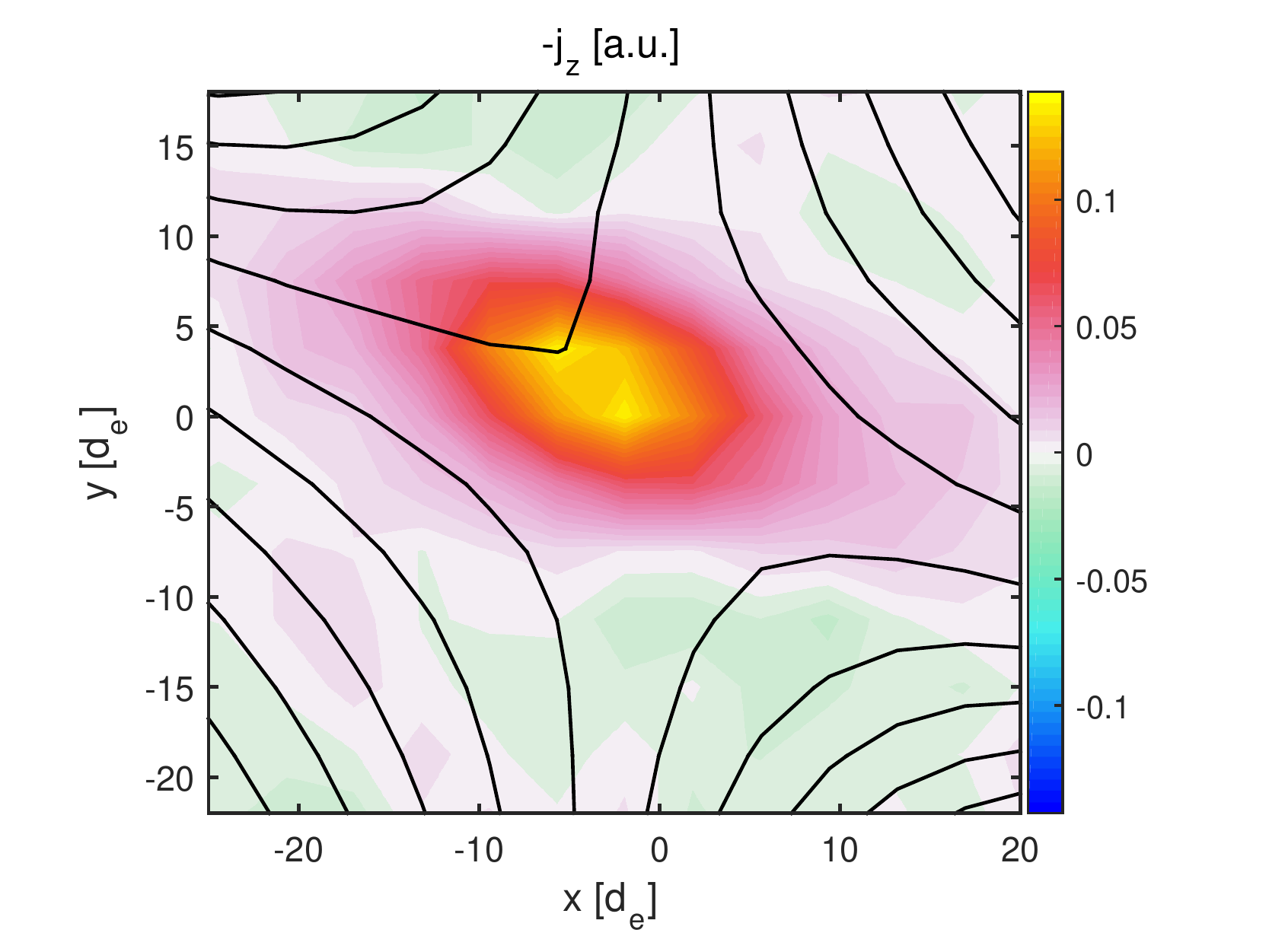}
\includegraphics[clip,width=0.48\textwidth,trim=1cm 0.3cm 1cm 1cm]{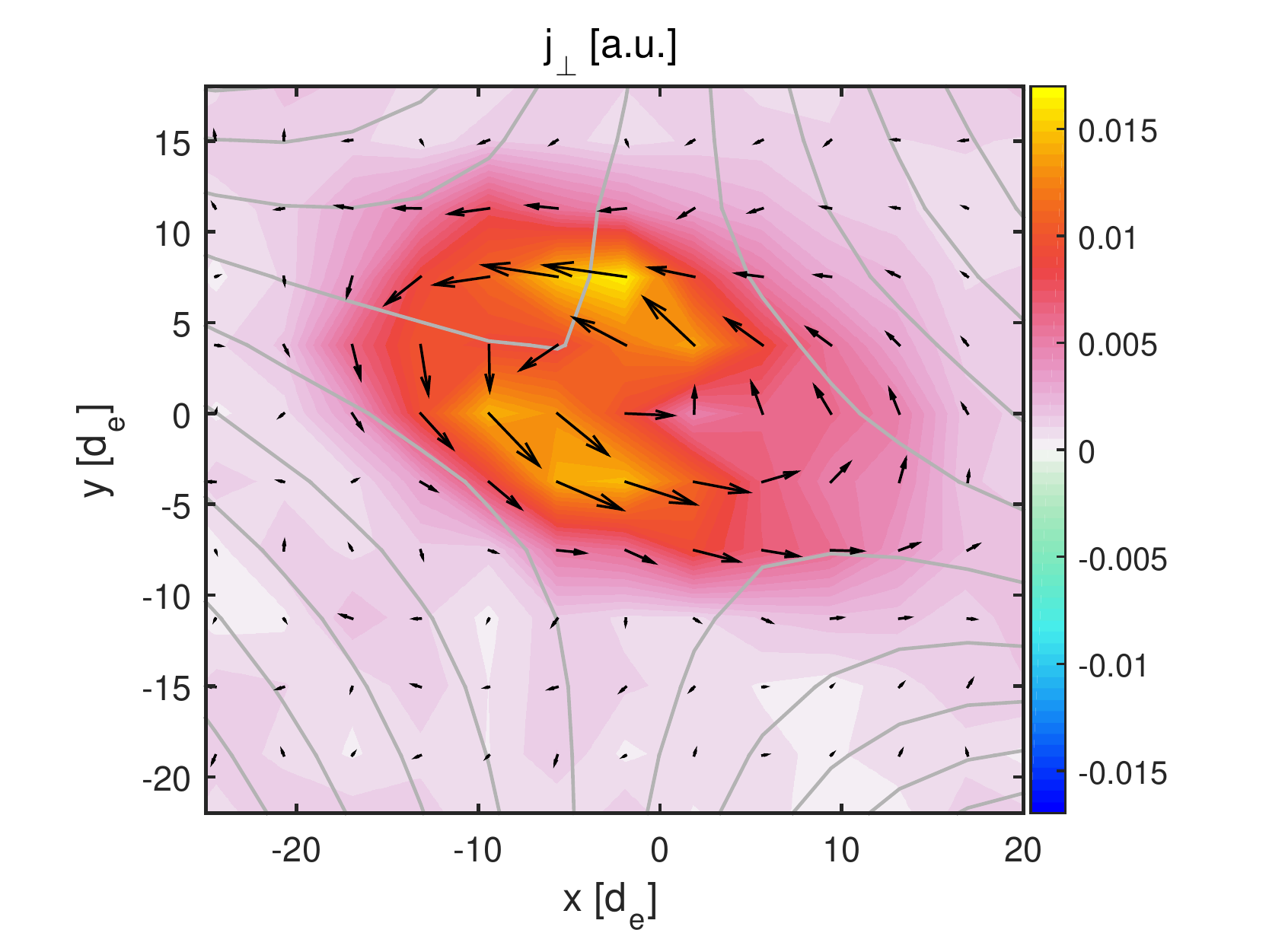}
\put(-430,175){\large \bf (a)}
\put(-200,175){\large \bf (b)}
\put(-265,175){\large \bf -$\mathrm{j_z}$}
\put(-45,175){\large \bf $|\mathrm{j_{\perp}}|$}
\caption{Experimental results at the time when axial plasma current is maximum. 2D cut of measurement data showing the color-coded axial current density $j_z$ 
(a). 
Vectors (arrow) and magnitude (color) of the in-plane current density (b). Field lines of the total perpendicular magnetic field 
are shown by black and gray lines in (a) and (b), respectively.}
\label{fig:exp-jz-jperp}
\end{figure}

In the \textsc{Vineta}.II experiments \cite{stechow2016} carried out in a linear device, magnetic reconnection is driven by an externally applied time varying magnetic field, which has a uniform and constant guide field component $B_{g}$ along $z$ (out-of-plane) and a figure-eight X-point field $\mathbf{B}_{\perp}(t)$ in the perpendicular $x$-$y$ plane (cf. cartoon in Fig. \ref{fig:exp-overview}). The perpendicular (in-plane) magnetic field is created by current-carrying conductors running parallel to the $z$-axis and is modulated in time to drive an inductive electric field $E_z$. Alternatively, an electrostatic field can be applied by biasing one axial boundary against the other without modulating in time  the perpendicular X-point magnetic field of the external currents. The characteristic experimental results on the electromagnetic fluctuations in an electron current sheet are similar in the two cases of electrostatic and inductive electric field drives.
An electron current sheet with a three-dimensional structure and a dominant axial component ($\mathbf{j}_{e}\approx j_{z}\mathbf{e}_z$)
forms in response to the external electric field and expands along the separatrices into an elongated shape as the electrons travel along the experiment axis. Fig. \ref{fig:exp-jz-jperp} shows a two-dimensional cut of the perpendicular plane at the point in time when the axial plasma current maximizes. The color plot in Fig.\ref{fig:exp-jz-jperp} (a) shows the axial current density, as reconstructed from local magnetic field measurements, in the $x$-$y$ plane in the normalized units of the simulations ($j_z$ by $n_0ev_{Ae}$, $x$ and $y$ by $d_e$).
 The un-normalized peak value of $j_z$ is 40\,kA/m$^2$. 
A detailed spatial analysis of the current sheet shows that while the electrons far from the X-point generally follow the downstream (left-right) separatrices, 
its central region close to the X-point tends to align with the axis between the conductors (top left and bottom right) as the plasma current rises and distorts the local magnetic field. The in-plane current, shown in Fig.\ref{fig:exp-jz-jperp} (b), develops a divergence-free vortex structure with its maximum located in the steepest gradients of axial current density, plasma density and electron temperature.

The time scale $1/f_r$ at which reconnection is driven (and during which the current sheet dynamics evolves) lies between the ion and electron cyclotron times, i.e. $1/f_{ce,g}\approx 1\,\mathrm{ns}\ll 1/f_r\approx 10\,\mathrm{\mu s}\ll 1/f_{ci,g}\approx100\,\mathrm{\mu s}$. 
The spatial scales $\delta$ lie between electron and ion scales, i.e. $\rho_{e,g} \approx 0.5\,\mathrm{mm} < d_e \approx 2.4\,\mathrm{mm} < \delta \approx 10\, \mathrm{mm}< \rho_{i,g}\approx 19.2 \mathrm{mm}\ll d_i \approx 0.6\,\mathrm{m}$. Here the electron and ion cyclotron frequencies ($f_{ce,g}$ and $f_{ci,g}$) and gyro-radii ($\rho_{e,g}$ and $\rho_{i,g}$) are based on the guide magnetic field $B_g$. At these scales, the EMHD model  
 should give an approximate phyiscal representation of the experimental system.

\begin{figure}[h]
\includegraphics[width=0.327\textwidth]{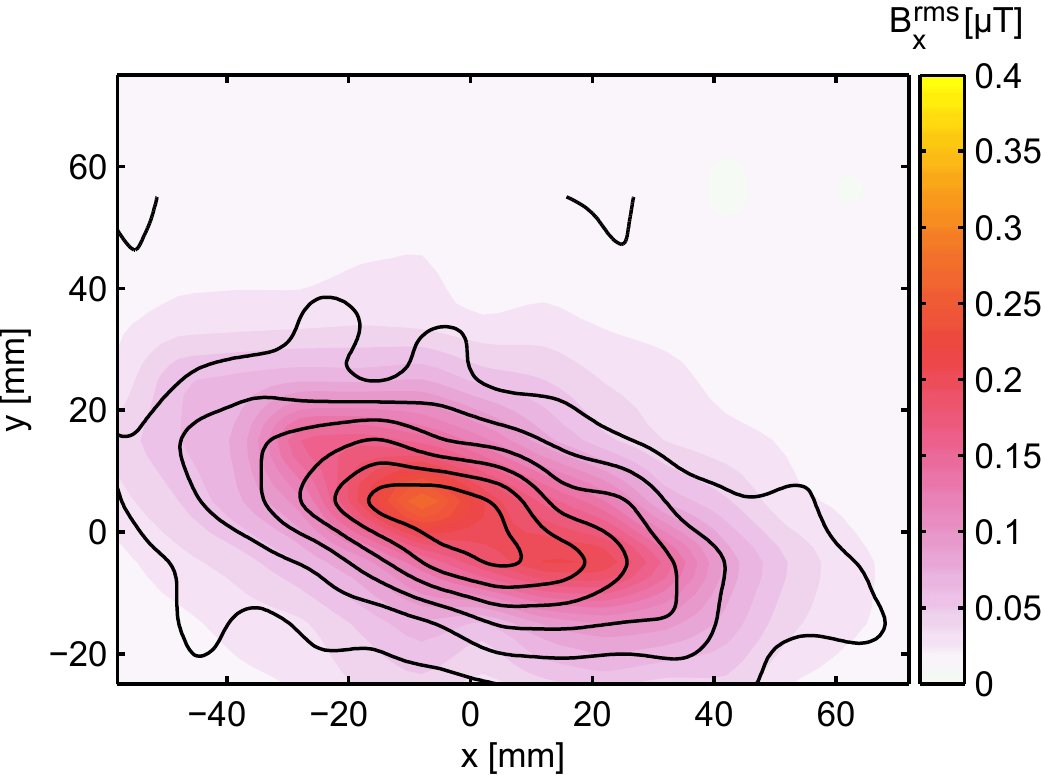}
\includegraphics[width=0.29\textwidth]{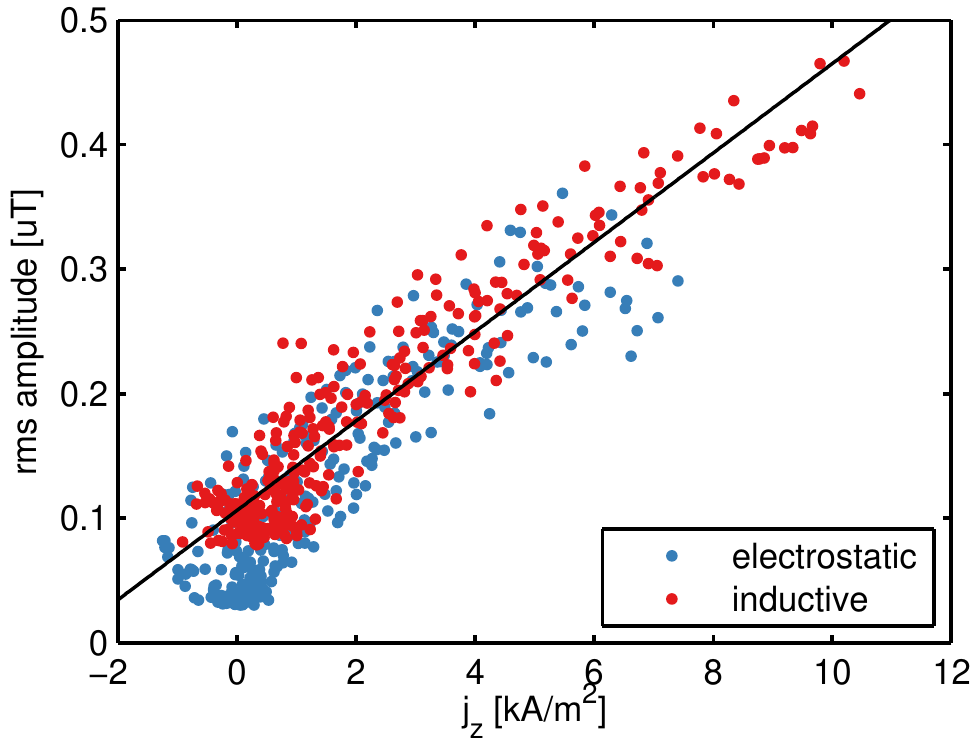}
\includegraphics[width=0.28\textwidth]{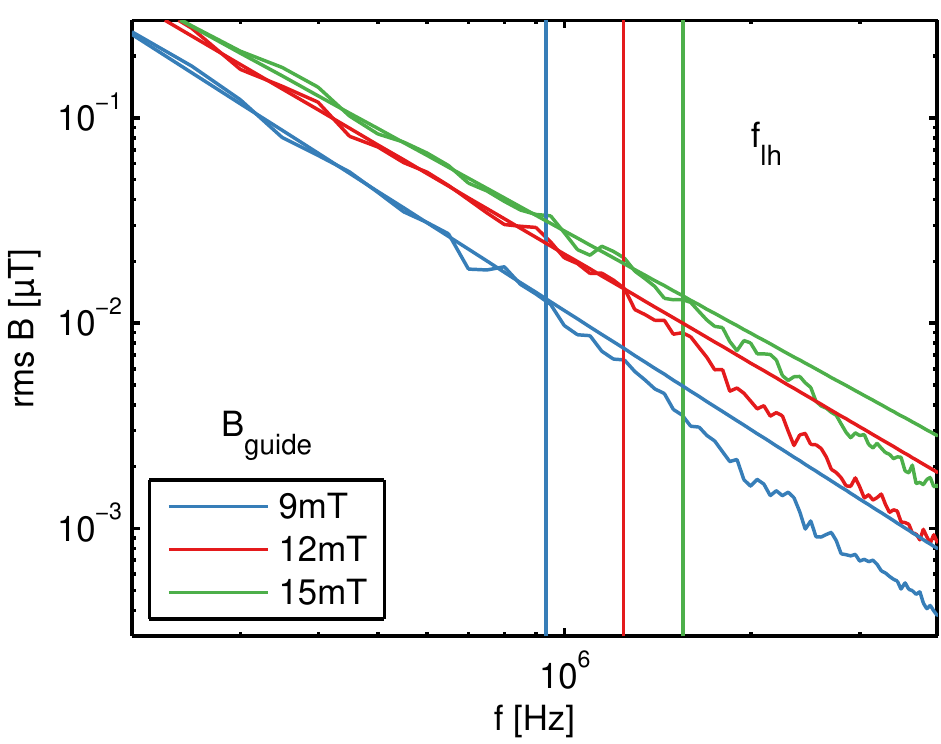}
\caption{Experimental results.
Left: color-coded rms amplitude of magnetic fluctuation $B_x^{rms}$ with contours of the axial current density.
Center: $B_x^{rms}$ vs. axial current density $j_z$ across space and time for different drive modes.
Right: amplitude spectra of $B_x$ at current sheet center (X-line) for different guide field strengths. For each value of the guide field, a vertical line crossing the frequency axis at the lower hybrid frequency in that guide field is drawn in the right figure. 
}
\label{fig:exp-flucs}
\end{figure}

A further key observation is the occurrence of high-frequency magnetic fluctuations throughout the electron current sheet. These are indicators of small-scale instabilities which are of interest especially in collisionless astrophysical systems where the  observed high reconnection rates require a non-resistive source of energy dissipation. While the typical fluctuation amplitudes in the experiments are by far too small to significantly contribute to the reconnection rate, their reproducibility allows for a detailed characterization of their spectral and propagation properties. Fig. \ref{fig:exp-flucs} summarizes the experimental results on fluctuations:
They occur throughout the current sheet (left, black contours), and their amplitudes (color-coded) maximize at the current sheet center. The fluctuation amplitude is observed to correlate well, in fact scaling almost linearly, with the local axial current density $j_z$ (middle). This is true not only across the measurement plane for a given point in time, but also through time as the current density rises and falls in response to the applied fields. The fluctuation spectra (right, shown for several guide field strengths) are broadband and their RMS amplitude $B_x^{rms}$  fall off with frequency $f$ according to a power law $B_x^{rms}(f)\sim f^{-\alpha}$. The spectra further reveal two guide field effects: First, the total fluctuation amplitude increases with rising guide field. This aspect can be traced back to an increased local current density as the current sheet is constrained to a smaller area at high guide fields. Second, a weak spectral break is observed near the lower hybrid frequency
(from $\alpha$=1.4-1.6 to $\alpha$=2.4) which shifts with the guide field strength, indicating an involvement of ions in shaping the fluctuation dynamics.
Since the EMHD model by definition does not account for ion motion, the latter observation is not expected to be reproduced by the simulations. Instead, this work focuses on possible sources of the broadband, electron scale ($f>f_{lh}$) fluctuations.

\section{Simulation setup \label{sec:simu_setup}}
\begin{figure}[h]
\includegraphics[width=0.5\textwidth]{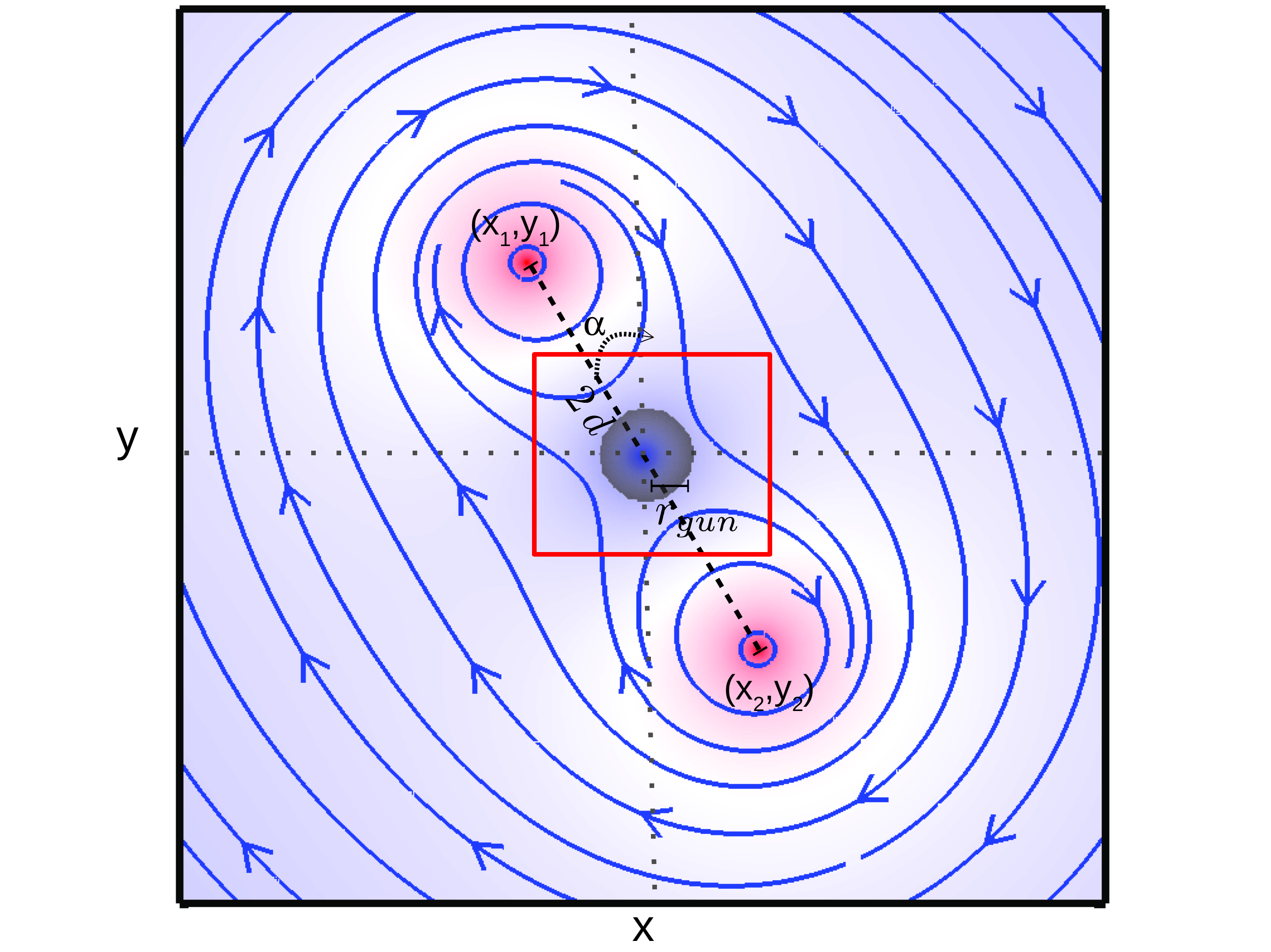}
\caption{Color-coded magnitude (red: high and blue: low) and field lines (blue) of the perpendicular external magnetic field $\mathbf{B}_{\perp}^{ext}$ produced by  two long wires at the labeled positions ($x_1,y_1$) and ($x_2,y_2$) carrying current in the negative z-direction. The shaded area at the center of the simulation box (red square) represents the extent of the external electric field $E_z^{ext}$. }
\label{fig:x-point}
\end{figure}

The simulations are initialized with a uniform, motionless plasma embedded in an externally imposed magnetic field $\mathbf{B}^{ext}=\mathbf{B}_{\perp}^{ext}+\mathbf{B}^{ext}_z$. The perpendicular magnetic field is created by two infinitely long wires separated by a distance $2\,d$ and carrying a current $I_0$ along the negative $z$-direction (see Fig. \ref{fig:x-point}):
\begin{equation}
\mathbf{B}^{ext}_\perp(\mathbf{r}) =
\frac{\mu_0I_0}{4\pi}
\left(\int_{-\infty}^{\infty} \frac{d\mathbf{z} \times (\mathbf{r}-\mathbf{r_1})}{|\mathbf{r}-\mathbf{r_1}|^3} + 
\int_{-\infty}^{\infty} \frac{d\mathbf{z} \times (\mathbf{r}-\mathbf{r_2})}{|\mathbf{r}-\mathbf{r_2}|^3}
\right),
\end{equation}
where $\mathbf{r}_{1,2}$:($x_{1,2},y_{1,2}$) are the wire positions. To match the experiment geometry, we take $2d=30\,$cm and the wire positions $\mathbf{r}_{1,2}$ to be rotated by $\alpha =30^\circ$ with respect to the $y$ axis. The guide field $B_g$ is set to $B_g=B_z^{ext}=15\,$mT and the current in the wires to $I_0=2$\,kA. 

In the experiments, the plasma current is extracted from a localized electron source (the plasma gun) with a radius $r_{gun}=6$ mm by the large scale electric field $E_z$, which can be either inductive or electrostatic. During operation, a complex force balance establishes, setting up additional axial electrostatic fields due to localized potential sheaths that establish at the experiment's axial boundaries.
The resulting effective electric field $E_z = -\partial A_z/\partial t-\partial\Phi/\partial z$ (where $A_z$ is the $z$-component of the magnetic vector potential and $\Phi$ is the scalar electrostatic potential)
that accelerates electrons roughly corresponds to the plasma gun shape and is approximated in the simulations by
\begin{equation}
\mathbf{E}_z^{ext}=-E_{z0}\exp\left(-\frac{x^2+y^2}{2r_{gun}^2}\right)\,\hat{z},
\end{equation} 
with $E_{z0}=10\,$V/m. This choice allows a reproduction of the experimentally observed current sheet evolution without requiring complex wall boundary conditions or detailed modeling of the plasma source.
 Though $E_z^{ext}$ is constant in time, corresponding more to the electrostatic  current drive experiments, the results apply to the inductive current drive experiments as well, because the time scales of the electron shear flow instabilities and resulting fluctuations  are much shorter than those of the inductive current drive.


The simulation box size $10\times 10\times 10\,$cm$^{-3}$ corresponds to the experimental measurement area. Note that the parallel conductors are outside the simulation domain (see Fig. \ref{fig:x-point}). The grid resolution in each direction is 1.35\,mm.  Simulations are run for $3.6\,\mu$s in steps of 0.2\,ns, which covers $1.5\cdot 10^3/f_{ce,g}$ or $6/f_{LH,g}$, where $f_{LH,g}=\sqrt{f_{ce,g}f_{ci,g}}$ is the lower hybrid frequency for an Argon plasma in the guide magnetic field $B_g=B_z^{ext}$.
The simulation results are presented in normalized variables: Magnetic fields are normalized by the edge magnetic field produced by the experimental current sheet ($B_{edge}=0.8\,$mT), time by the inverse electron cyclotron frequency in this magnetic field ($\omega_{ce}^{-1}=(eB_{edge}/m_e)^{-1}=7.2\,$ns), distances by the electron inertial length ($d_e=2.7\,$mm) and the current density by $n_0ev_{Ae}=B_{edge}/\mu_0d_e$=235.8 kA/m$^2$.

\section{Results\label{sec:results}}
The three dimensional EMHD simulations of the formation of an electron current sheet and subsequent development of the electromagnetic fluctuations show that the evolution of the system remains two-dimensional throughout the simulation runs. No significant variations were observed in the $z$-direction. For this reason, we show our results in the perpendicular $x$-$y$ plane at  $z=0$. 
\subsection{Current sheet formation\label{sec:ecs}}

\begin{figure}
\includegraphics[width=0.5\textwidth]{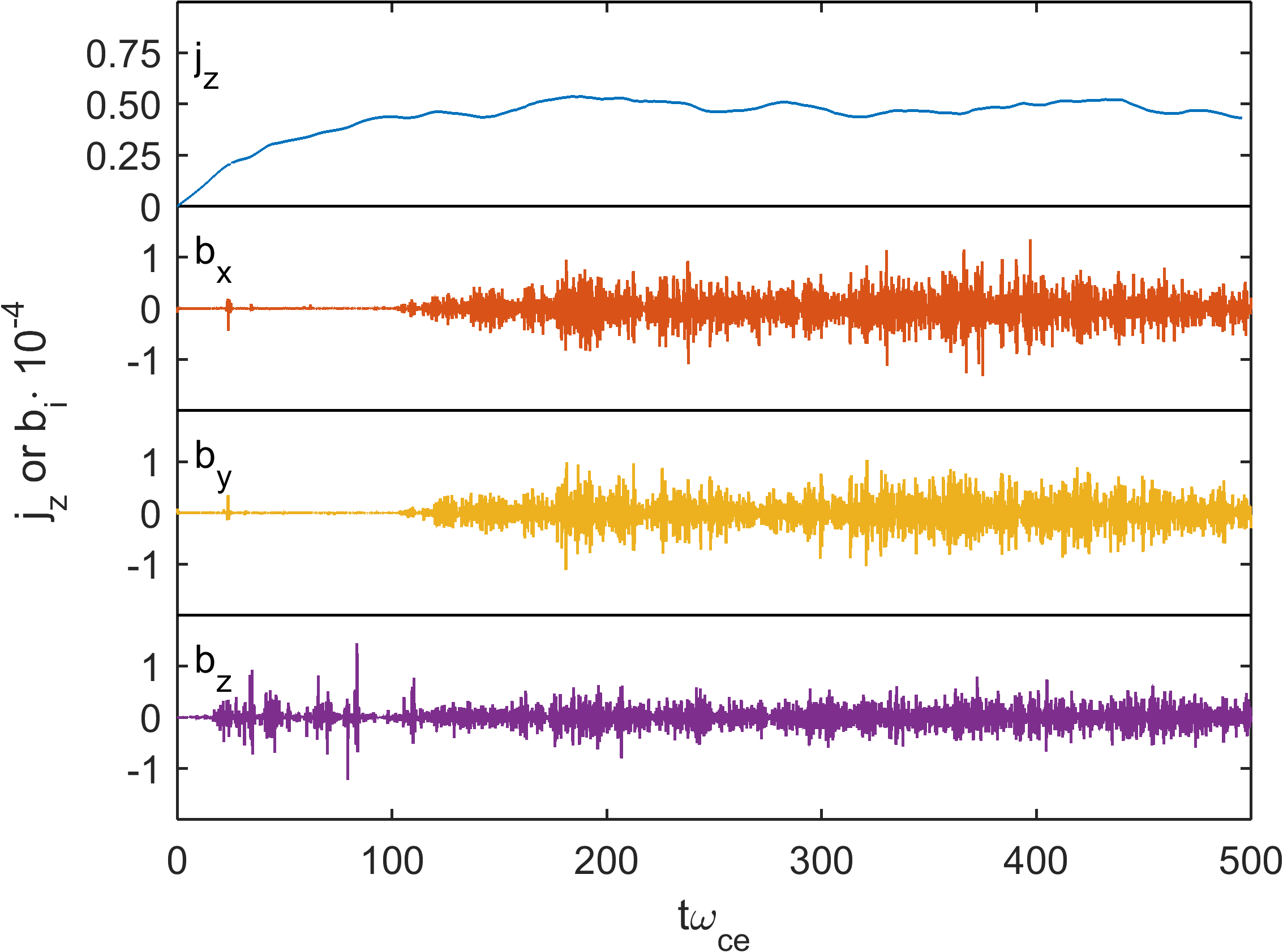}
\caption{Simulation results. Time evolution of the current density $j_z$ (top panel) and fluctuations in magnetic field components,  $b_i$, ($i=x, \, y, \, z$, bottom three panels), obtained by  high pass filtering ($f_{-3\,\mathrm{dB}}=f_{ce}$)  the plasma magnetic field near the X-line. }
\label{fig:jz_time}
\end{figure}

\begin{center}
\begin{figure}[h]
\includegraphics[clip,width=0.32\textwidth,trim=1cm 0.3cm 1cm 0.2cm]{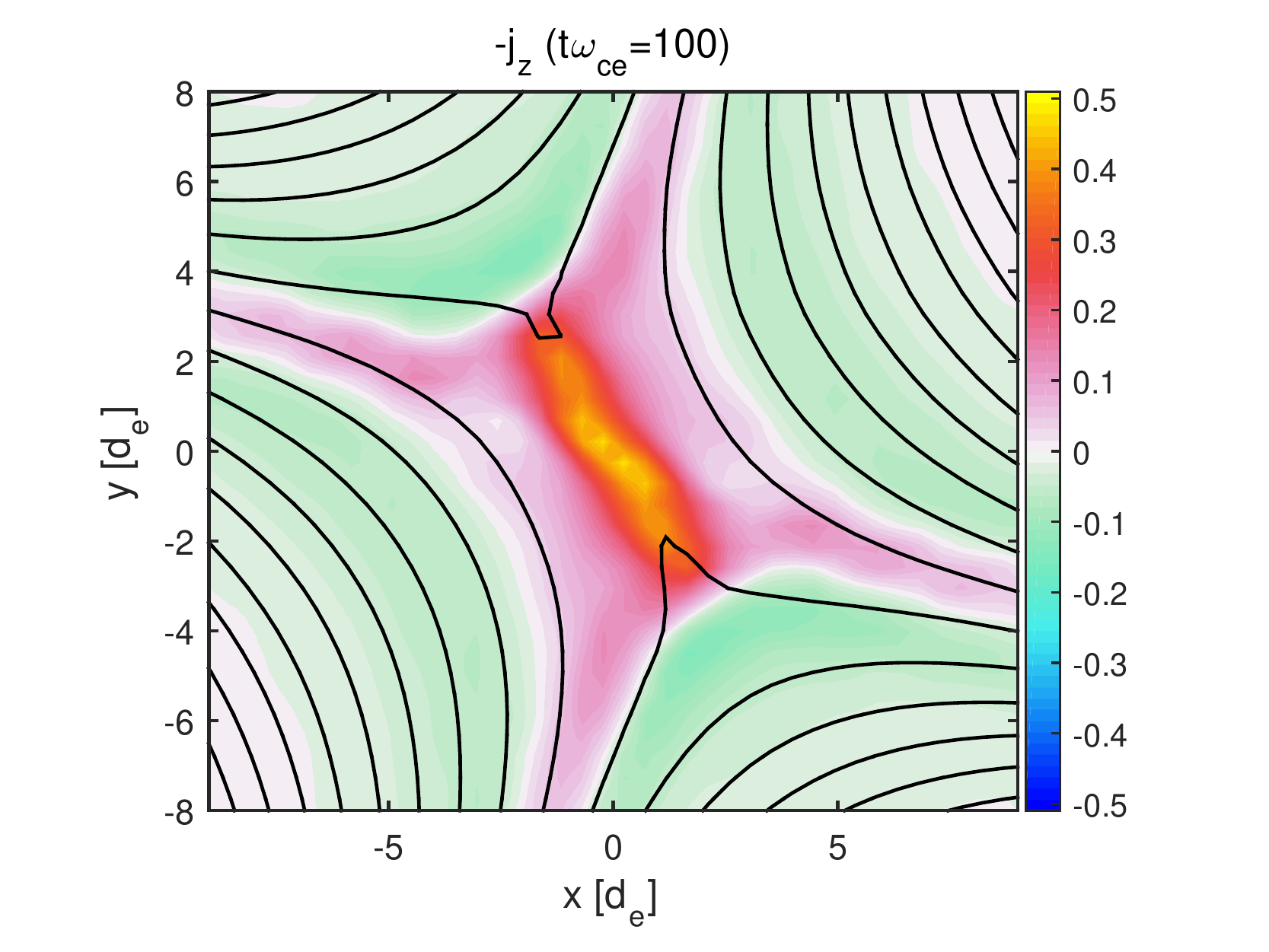}
\put(-135,118){(a)}
\includegraphics[clip,width=0.32\textwidth,trim=1cm 0.3cm 1cm 0.2cm]{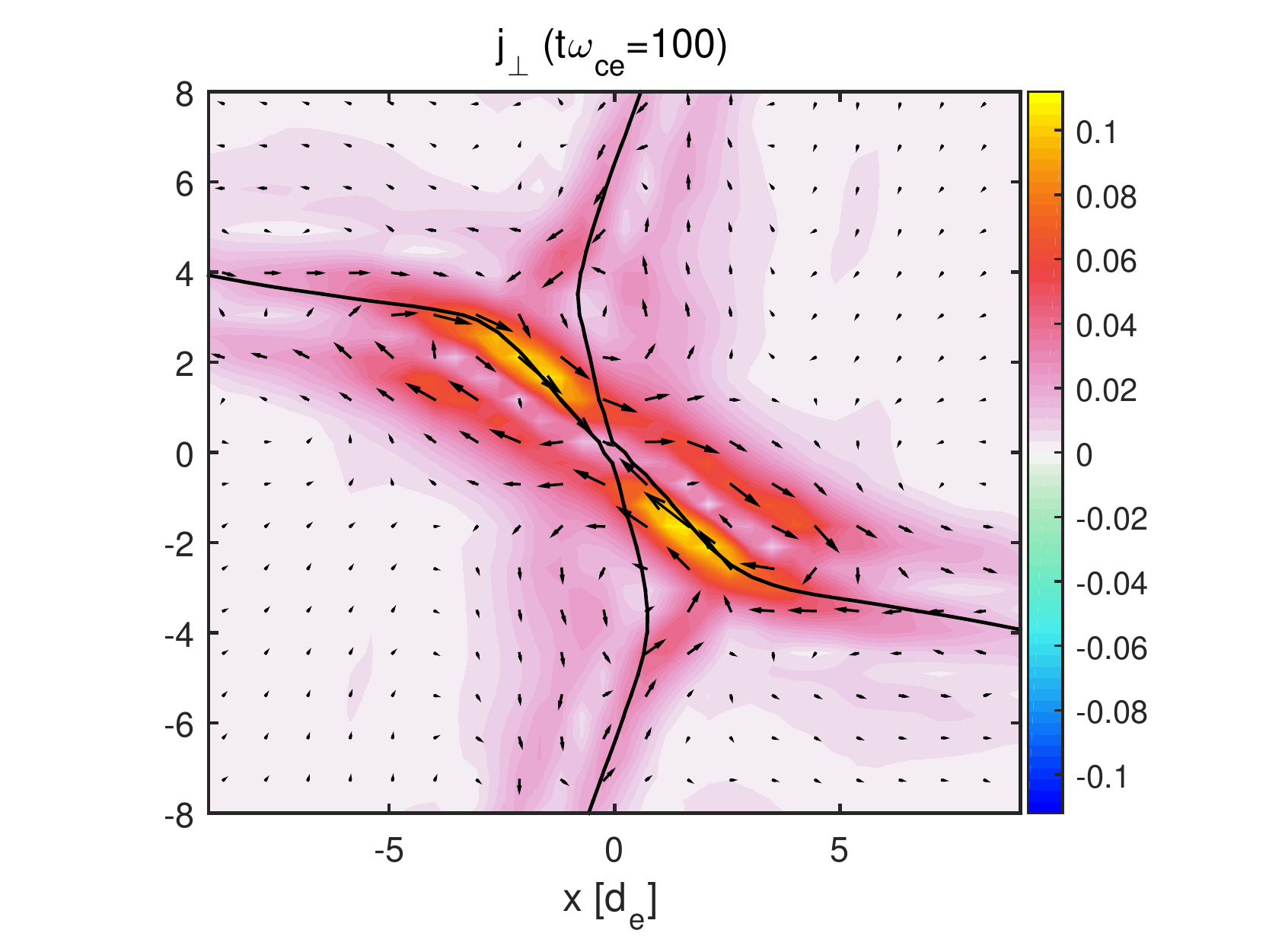}
\put(-135,118){(b)}
\includegraphics[clip,width=0.32\textwidth,trim=1cm 0.3cm 1cm 0.2cm]{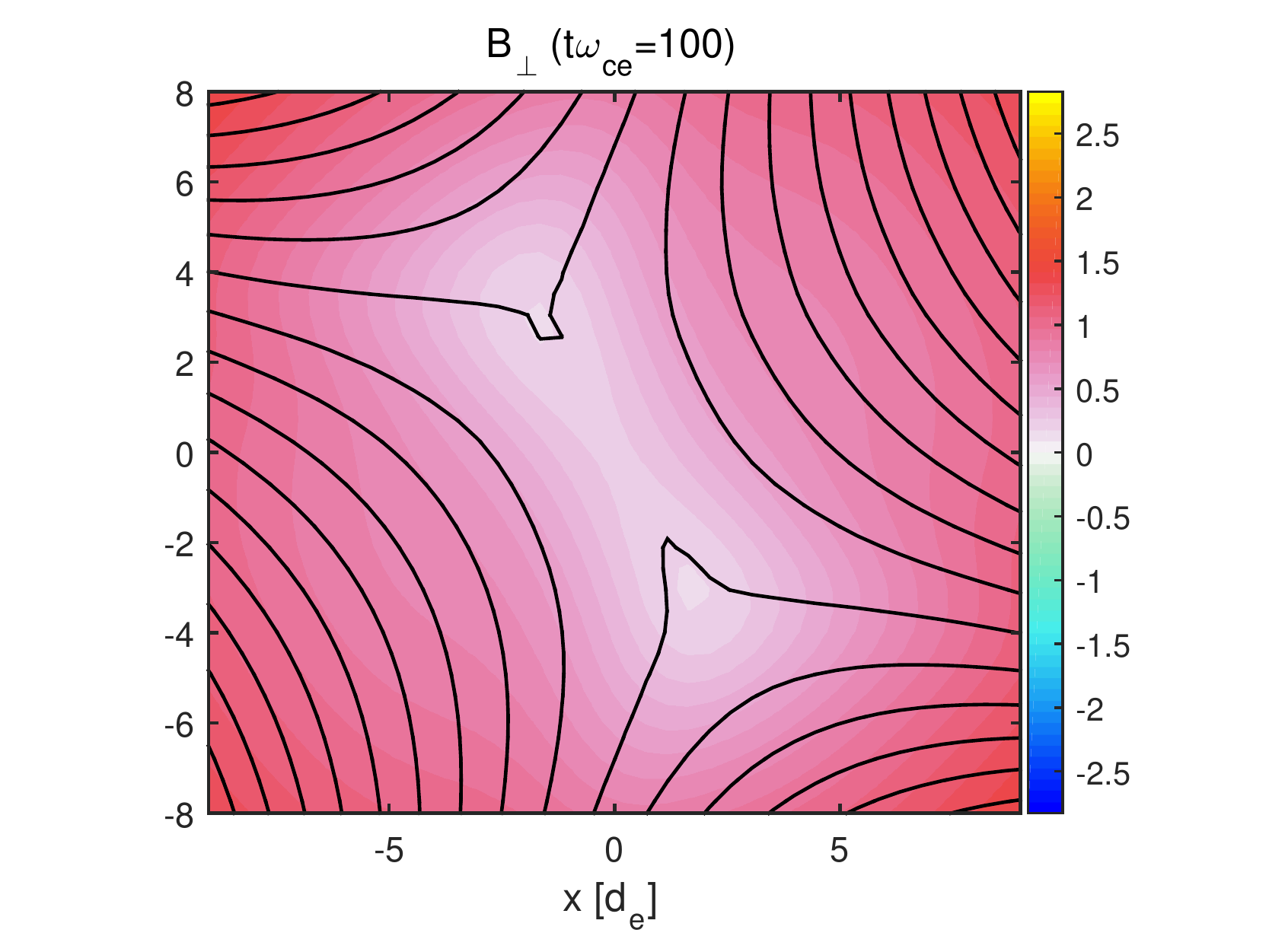}
\put(-135,118){(c)}
\caption{Simulation results. Color-coded axial (a) and in-plane (b) current density with field lines (black) of the total perpendicular magnetic field $\mathbf{B}_{\perp}=\mathbf{B}_{\perp}^{p}+\mathbf{B}_{\perp}^{ext}$, where $\mathbf{B}^p$ is the plasma magnetic field due to plasma currents. Arrows in (b)  point in the direction of the in-plane current density. Magnitude (color) of $\mathbf{B}_{\perp}$ and its field lines (black) in (c).
}
\label{fig:jz_slice}
\end{figure}
\end{center}

In response to the external electric field $E_z^{ext}$, the electron current density $j_z$ develops around the central X-point. In the vicinity of the X-point where the in-plane (perpendicular to the guide magnetic field) magnetic field is vanishingly small, electrons are accelerated by $E_z^{ext}$ along the $z$-direction.
Fig. \ref{fig:jz_time} shows that the current density $j_z$ at the X-point grows in time until $\omega_{ce}t \approx 100$ and then saturates around $j_z=0.5\, n_0ev_{Ae}\approx 118$ kA/m$^2$, which is of the same order of magnitude as the peak experimental current density (40 kA/m$^2$).  


The rising out-of-plane current density $j_z$ generates not only its own  magnetic field  creating a neutral sheet in the total perpendicular magnetic field $\mathbf{B}_{\perp}=\mathbf{B}_{\perp}^{p}+\mathbf{B}_{\perp}^{ext}$ ($\mathbf{B}^{p}$ is the plasma magnetic field due to plasma currents), shown in Fig. \ref{fig:jz_slice} (c), but also an inductive electric field $E_{ind}=-dA_z/dt$, where $\nabla^2 A_z = -\mu_0 j_z$. The induced electric field, by the nature of the Poisson equation, is broader than the current sheet. This field, which has the opposite sign of the externally applied and localized electric field, drives a return current which flows at the edges of the current sheet as shown in Fig. \ref{fig:jz_slice} (a). 

The development of plasma currents and electromagnetic fields quickly establish a force balance between in-plane electric and magnetic forces, i.e. $[\mathbf{E}+\mathbf{v}_e\times\mathbf{B}]_{\perp}\approx 0$.
Consequently the electron's in-plane drift velocity is primarily given by $\mathbf{v}_{e\perp}=\mathbf{E}_{\perp}\times \hat{z}/B_z$.
These in-plane electron drifts modify the initially ($\omega_{ce}t < 10$) circular cross-section of $j_z$, respectively stretching and pinching the current channel along the directions parallel and perpendicular to the line connecting the conductors. Simultaneously, $j_z$ extends towards the separatrices. The resulting structure of the out-of-plane current sheet at $\omega_{ce}t=100$ is shown in Fig. \ref{fig:jz_slice} (a).
Note that $j_z$ flows along the negative z-direction in the current sheet as well as in the four separatrices. This is because the net out-of-plane electric field $E_z^{net}=E_z+\hat{z}.(\mathbf{v}_{e\perp}\times\mathbf{B}_{\perp})$ seen by the electron fluid is in the negative z-direction in the current sheet and the four separatrices, accelerating electrons in the positive z-direction.   

\begin{center}
\begin{figure}[h]
\includegraphics[clip,width=0.35\textwidth,trim=1cm 0.3cm 0.1cm 0.2cm]{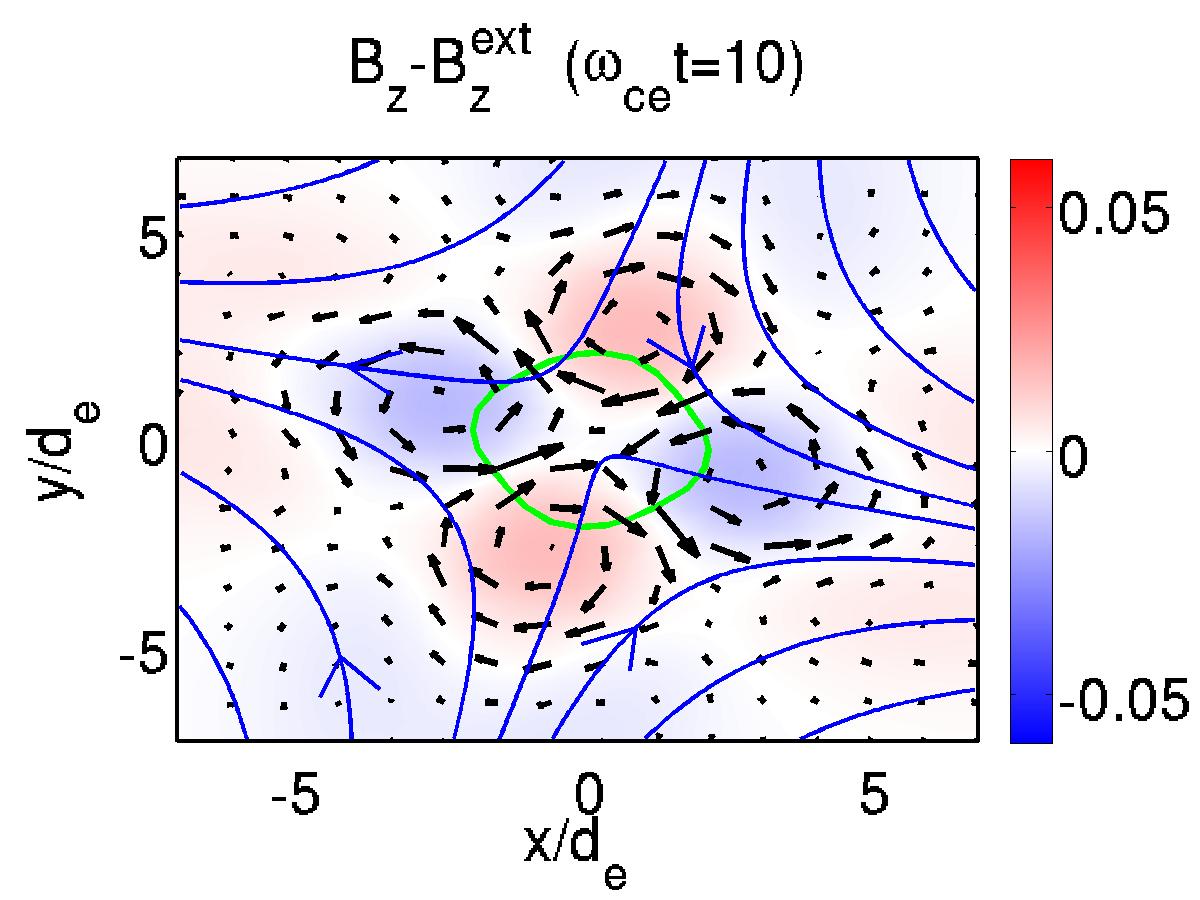}
\put(-145,110){(a)}
\includegraphics[clip,width=0.35\textwidth,trim=1cm 0.3cm 0.1cm 0.2cm]{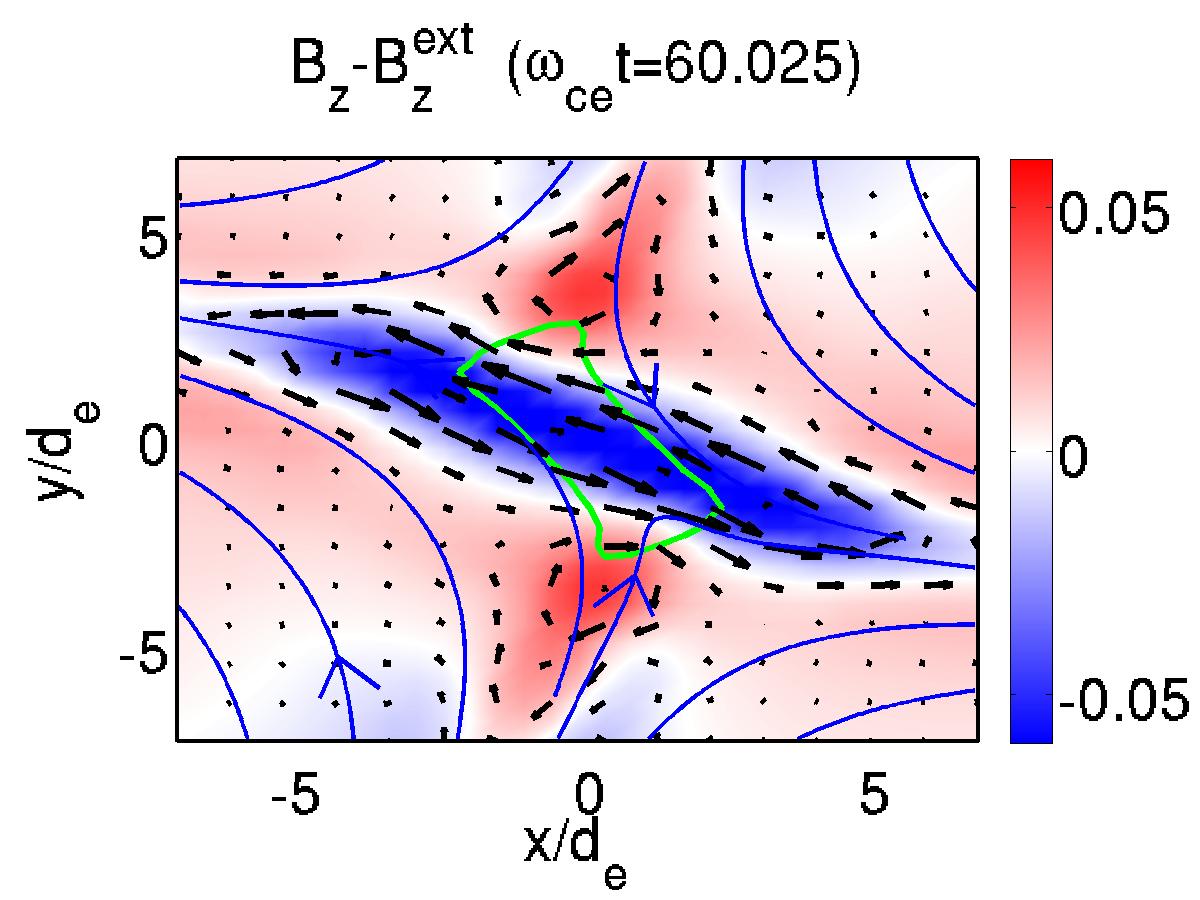}
\put(-145,110){(b)}
\includegraphics[clip,width=0.35\textwidth,trim=1cm 0.3cm 0.1cm 0.2cm]{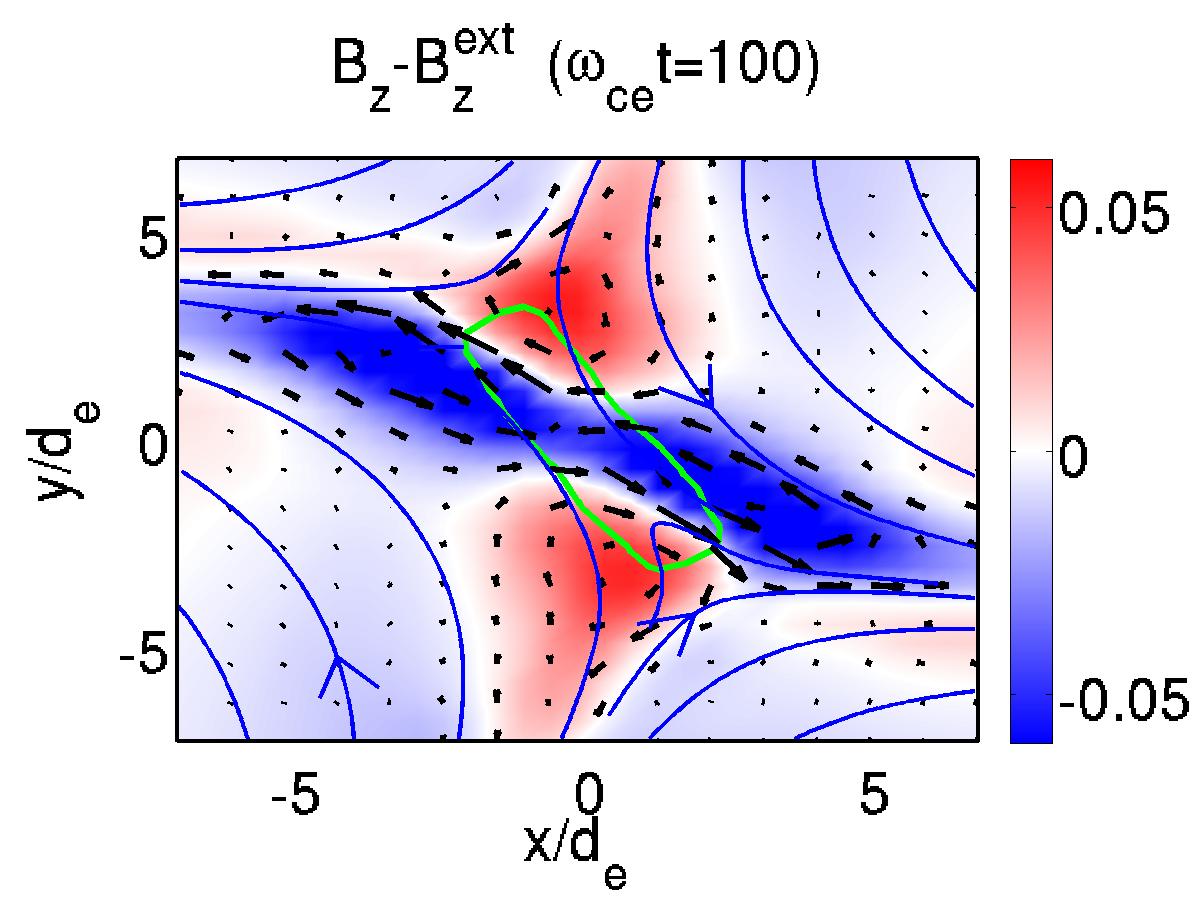}
\put(-145,110){(c)}
\caption{Simulation results. Out-of-plane component of the plasma magnetic field, $\hat{z}.(\mathbf{B}-\mathbf{B}^{ext})$ (color), at three different times with projection of magnetic field lines (blue) and electron flow vectors (arrow) in an x-y plane, (a)-(c). Green curve is the contour of the axial current density at $j_z=0.42\, j_z^{max}$.
}
\label{fig:bz_slice}
\end{figure}
\end{center}

The in-plane electron velocity  develops a shear flow structure. 
The magnitude of the in-plane current density ($\mathbf{j}_{\perp}=-\mathbf{v}_{e\perp}$) varies in the $x$-$y$ plane as shown in Fig. \ref{fig:jz_slice} (b). A similar structure of the in-plane currents was also observed in the experiments (see Fig. \ref{fig:exp-jz-jperp}b).
The development of the in-plane shear flow structure in the simulations from an initial time $\omega_{ce}t=10$ to $\omega_{ce}t=100$ is shown in Fig. \ref{fig:bz_slice} (a)-(c). At $\omega_{ce}t=10$, the out-of-plane component of the plasma magnetic field, $\hat{z}.(\mathbf{B}-\mathbf{B}^{ext})$, has a quadrupole structure consistent with the initial in-plane flows which pinch and stretch the current sheet. By $\omega_{ce}t \approx 60$, the left and right poles (or electron vortices) of the quadrupole have merged into each other pushing away the top and bottom poles.  The single vortex formed by the merging 
breaks again into two vortices as shown at $\omega_{ce}t=100$ in Fig. \ref{fig:bz_slice} (c). The pinching and stretching of the current sheet can also be seen in Fig. \ref{fig:bz_slice}. Note that the in-plane shear flow is aligned with the left-right separatrix pair and not with the current sheet itself. 

The merging and breaking of the electron vortices  are enabled by an electron shear flow instability, namely electron Kelvin-Helmholtz instability, triggered by electron inertia. In order to ensure the role of inertia, we carried out EMHD simulations of the same setup but without electron inertial terms. We found that the axial current density at the X-point continues to grow without saturating. Also the shear flow structures of the out-of-plane and in-plane flows, as shown in Figs. \ref{fig:jz_slice} and \ref{fig:bz_slice}, did not develop.

\subsection{Electromagnetic fluctuations in the current sheet\label{sec:fluctuation}}
High frequency electromagnetic fluctuations develop in the current sheet. Fig. \ref{fig:jz_time} shows 
the high frequency fluctuations in magnetic field components $b_j \,\,\,(j=x,y,z)$ near the X-line, obtained by high pass filtering ($f_{-3\mathrm{dB}}=f_{ce}$) the plasma magnetic field. The fluctuations in the $x$ and $y$ components begin to develop at around $\omega_{ce}t=100$, which coincides approximately with the saturation of the axial current density. The fluctuations in the $z$-component, on the other hand, develops earlier (see Fig. \ref{fig:jz_time}). Although the amplitude of $b_z$-fluctuations is relatively larger than the other components in the beginning,  the amplitudes of all three components at later times, $b_j/B_{edge}\sim 10^{-4}$, are of the same order of magnitude  in the selected frequency range.
Since the simulations do not include collisional resistivity terms, the question arises why the electron current saturates. The coincidence of the onset of in-plane fluctuations and the saturation of $j_z$ at $\omega_{ce}t\approx 100$ indicates that the electromagnetic fluctuations might provide some type of anomalous dissipation that balances the electric field at the X-point. 
Further investigations confirming this hypothesis will be presented elsewhere. 

\begin{center}
\begin{figure}
\includegraphics[clip,width=0.31\textwidth]{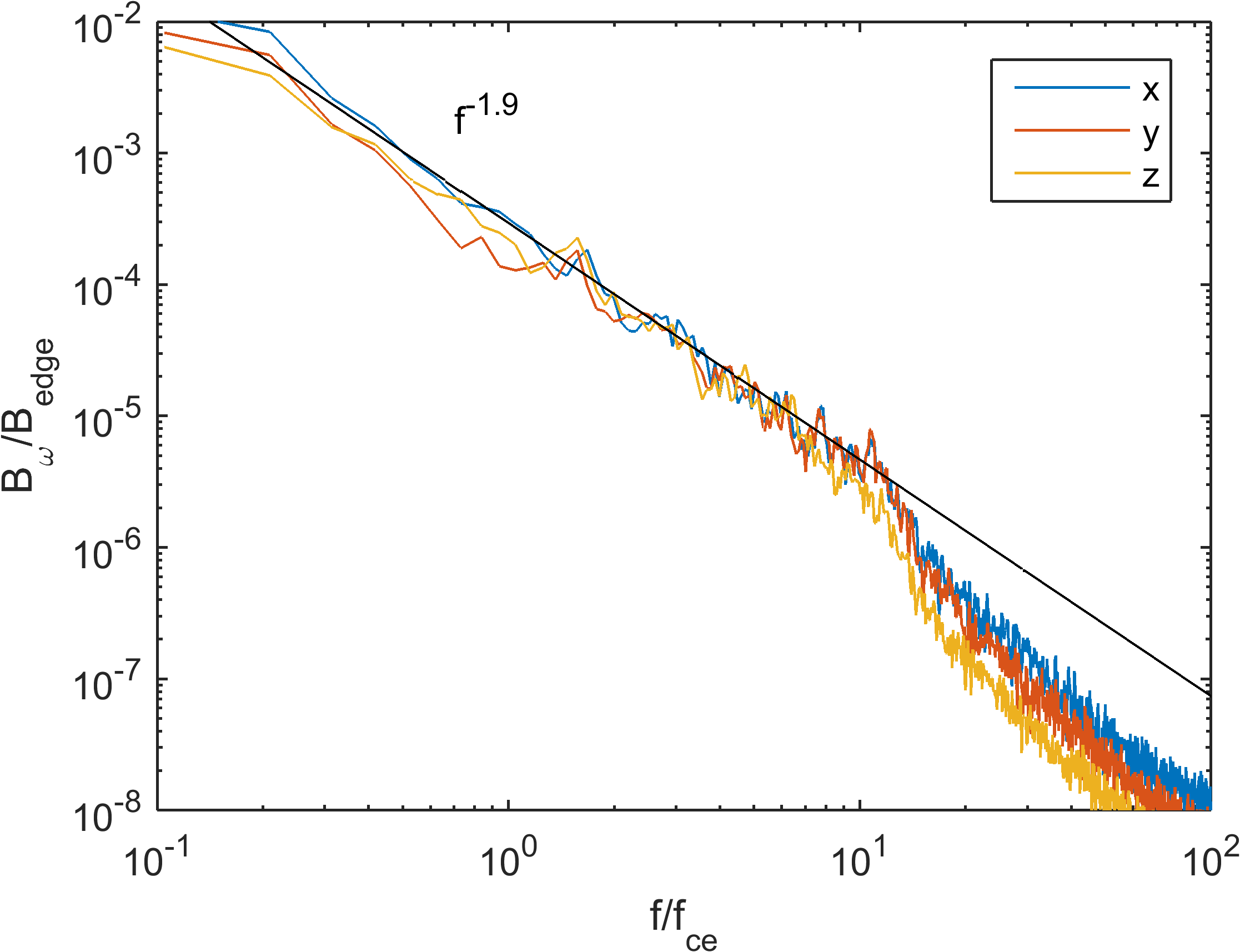}
\put(-140,120){(a)}
\includegraphics[clip,width=0.34\textwidth]{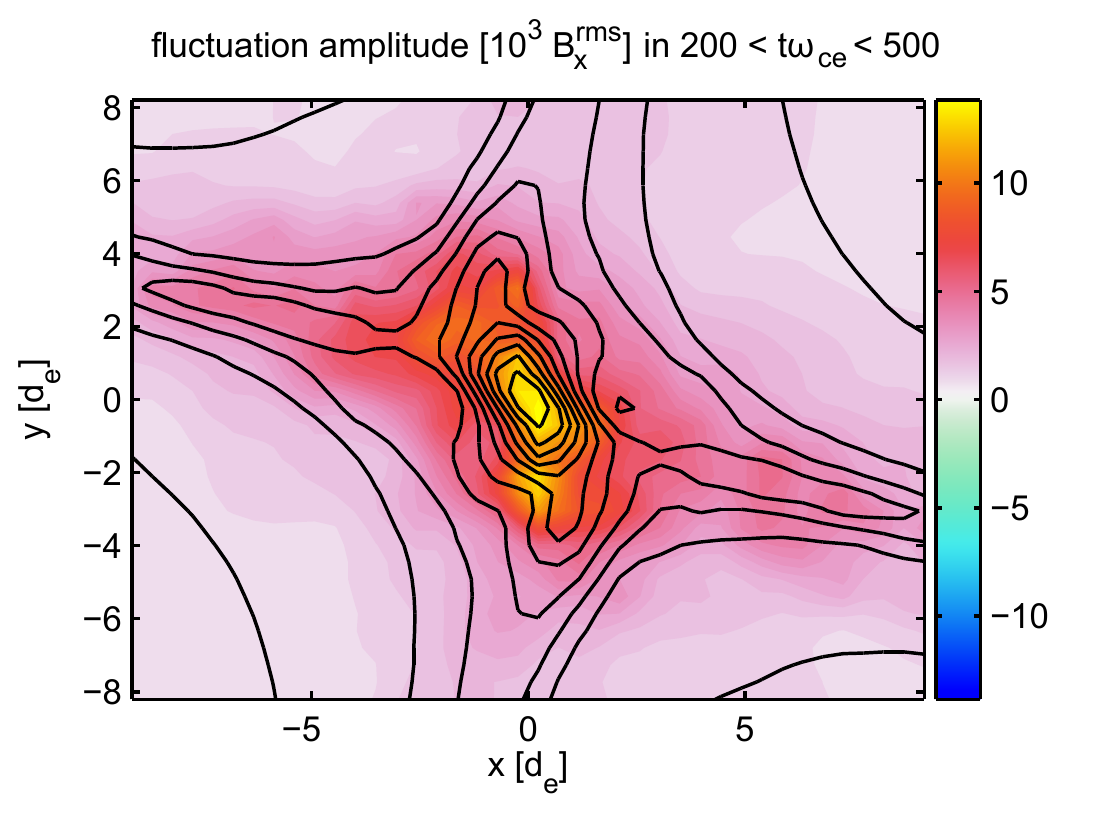}
\put(-140,120){(b)}
\includegraphics[clip,width=0.33\textwidth]{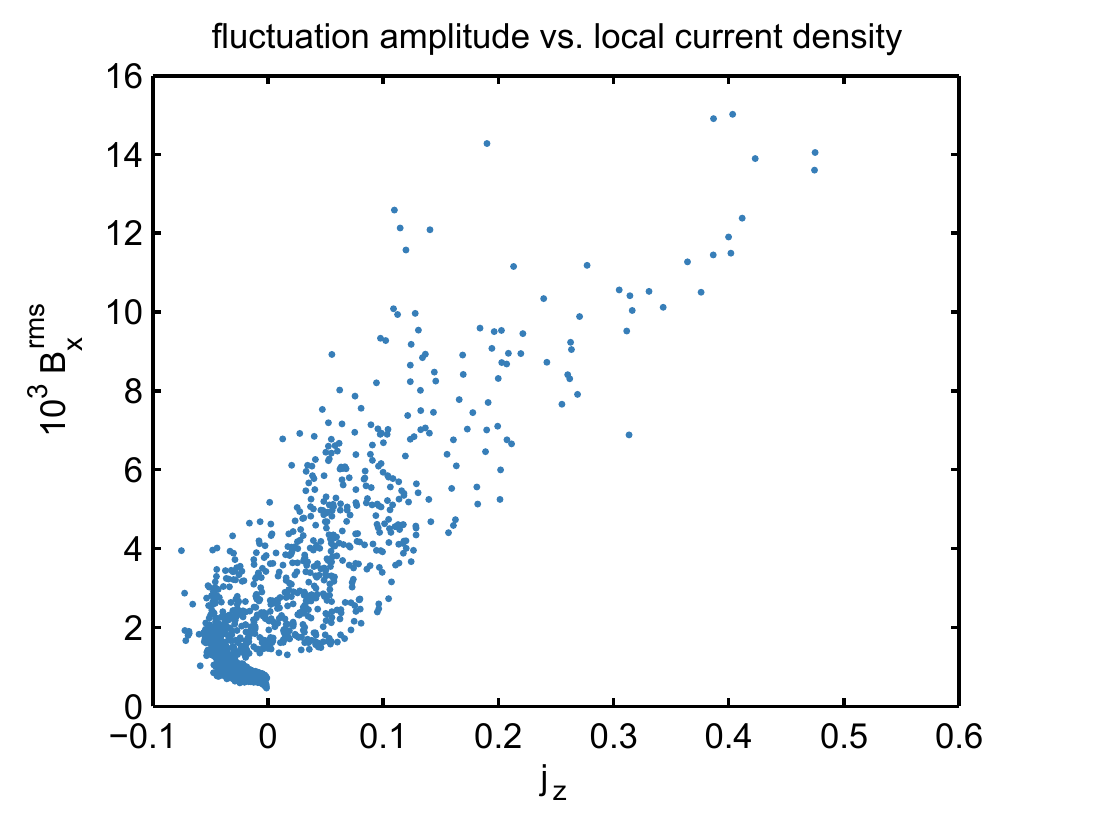}
\put(-140,120){(c)}
\caption{Simulation results. (a) Amplitude spectra of the $x$, $y$ and $z$ components of the plasma magnetic field $\mathbf{B}^{p}=\mathbf{B}-\mathbf{B}^{ext}$ at the X-point. (b) Root-mean-square (RMS) fluctuation amplitude (color) and contours of axial current density. (c) Scatter plot of  RMS fluctuations amplitudes with the axial current density over space. The Fourier (in a) and RMS (in b and c) amplitudes were obtained from the magnetic field data in the time interval  $200 <\omega_{ce}t<500$.
}
\label{fig:spectra}
\end{figure}
\end{center}

Fig. \ref{fig:spectra}(a) shows amplitude spectra of the plasma magnetic field components at the X-line during the time period in which the fluctuations are fully developed, i.e. $200 <\omega_{ce}t<500$. The spectra show broadband, possibly turbulent, behavior exhibiting a clear power law with a spectral index of $\alpha\approx1.9$ across two decades of frequencies ($f/f_{ce}=0.1-10$). Above $f=10\,f_{ce}\approx 0.5f_{ce,g}$, the spectra becomes steeper. The steepening of the spectra above a frequency of the order of electron cyclotron frequency has been observed in the solar wind turbulence as well \cite{alexandrova2009}.
All three magnetic field components have similar amplitudes and the same spectral index in the frequency range $f/f_{ce}=0.1-10$. The experimental spectra show broadband power law behavior over a wider frequency range, extending below the lower hybrid frequency $f_{LH,g}$ in the guide magnetic field. The part of the spectrum at or below $f_{LH,g}$, however,  can not be studied  in our simulations since ion motion is not included in the model.
On the other hand, the spectral index $\alpha \approx 1.9$ obtained in the simulations is comparable to the experimental value $\alpha \approx 2.4$ for the high frequency part ($f > f_{LH,g}$) of the spectrum.

Root-mean-square (RMS) values of plasma magnetic field components  are calculated from the simulation data using the same procedure as in the experiments, i.e.
\begin{equation}
{B}_j^{rms}(\mathbf{r}) = C\int_{f_{min}}^{f_{max}}
\left[ \int W(t) B_j^{p}(\mathbf{r},t)\exp(-i\,2\pi f t)\, dt\right]df, \,\,\,\,\,\,\,\, j=x,\, y,\, z
\end{equation}
\\
where the integral in the square brackets is  Fourier transform of $W(t)B_j^{p}(\mathbf{r},t)$,  $W(t)$ is a Hanning window and $C$ is the proper normalization factor to obtain the RMS magnetic field amplitude.
The results of this calculation over the time span $\omega_{ce}t=200-500$, with $f_{min}=0.05\,f_{ce}$ and $f_{max}=1.5\,f_{ce}$, yield a  spatial distribution of the fluctuation amplitude shown in Fig. \ref{fig:spectra} (b) together with contours of $j_z$. Analogous to the experiment, it is apparent that the fluctuations correlate well with the local current density and peak at the center of the current sheet. Fig. \ref{fig:spectra} (c) further shows that there is a good correlation between the two quantities with a nearly linear relationship, as is the case in the experiment.

\section{Conclusion}
\label{sec:conclusion}
We have carried out three dimensional EMHD simulations of the formation of an electron current sheet and subsequent generation of electromagnetic fluctuations.
We chose plasma parameters and a magnetic field configuration similar to those of the \textsc{Vineta.II} guide field reconnection  experiment.
In spite of the simplifications of the EMHD model, the simulations revealed many results analogous to the experiments, indicating some common physics in the simulations and experiments. Similar to the experiments, the simulations show the formation of an electron current sheet and in-plane vortical flow structures with comparable spatial scales. 
However unlike the experiments, the vortical structure of the in-plane electron flow velocity in the simulations does not align with the current sheet. Note that the vortical structure in the simulations is due to the $E \times B$ drift of electrons only, but, in the experiments, dimagnetic drifts due to pressure gradient also contribute to the structure of the in-plane flows. The pressure gradients can also contribute to the thickness of the experimental current sheet which is approximately three times its value in the simulations. 

Electromagnetic turbulence within the current sheet is obtained, though concentrated around the electron cyclotron frequency, as is typical for the EMHD approach. The turbulence exhibits a power law, $B_f \propto f^{-\alpha}$,  with a spectral index $\alpha=1.9$. As in the experiments, RMS magnetic field fluctuation amplitudes correlate well with the local axial current density. 
Not all the characteristics of the experimentally observed turbulence could be reproduced in the simulations.  For example, 
the measured magnetic fluctuation amplitudes of the axial component are much smaller than the perpendicular ones, whereas the amplitudes in the simulation are similar in all three components. The spectral index $\alpha$ in the simulations differs from its measured value by approximately 20\%. These differences could be due to the absence of ion physics in our model.
We conjecture that the underlying basic instability mechanism responsible for the generation of turbulence in the experiments  is the same as in our simulations but with the ion effects.

\begin{figure}
\includegraphics[width=0.5\textwidth]{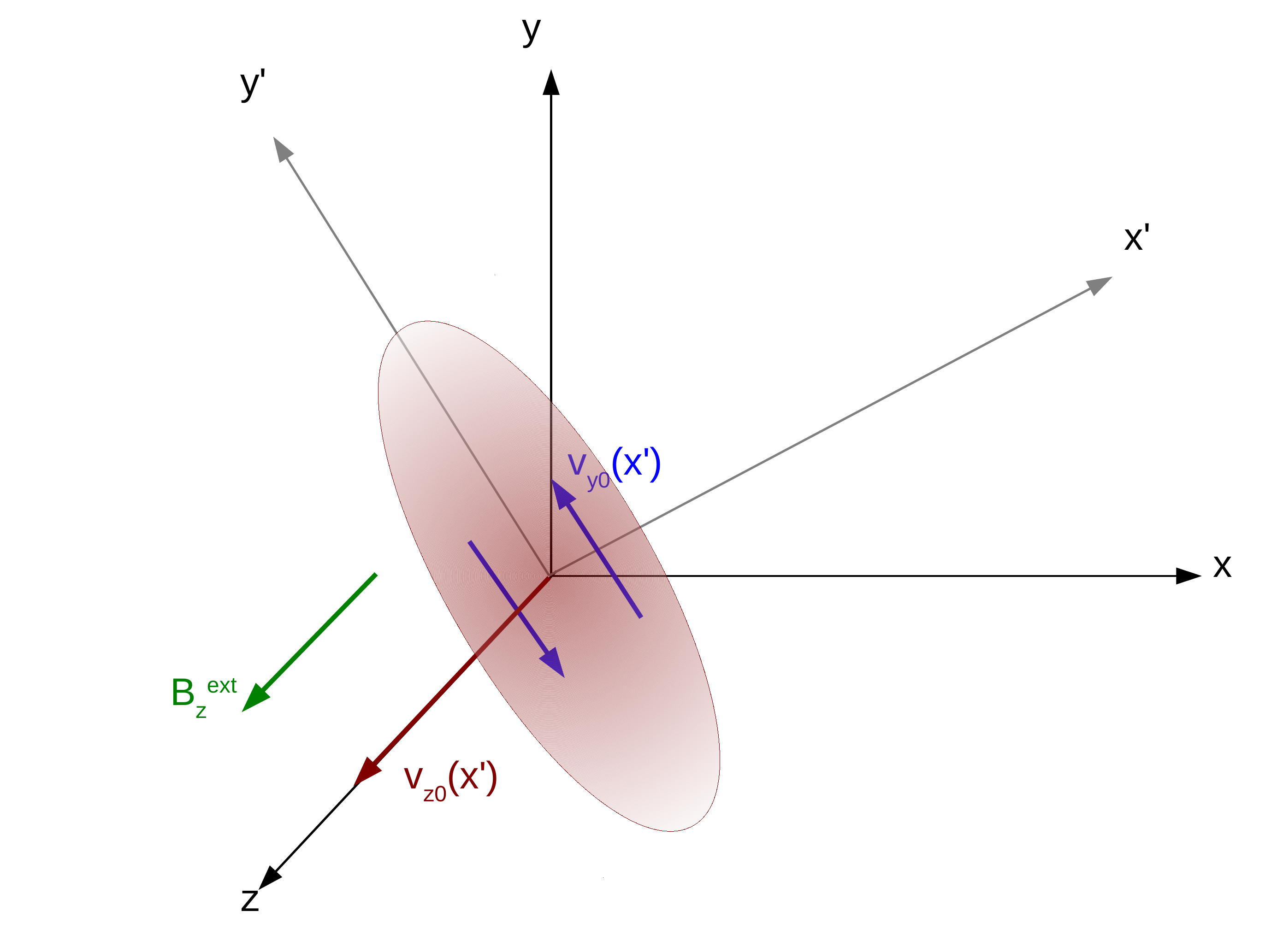}
\caption{\label{fig:two_shear_flow} Schematic of shear flows developed in the simulations. The ellipse represents the electron current sheet in x-y plane with electron current directed along $-z$ axis. $x'$ and $y'$ are the minor and major axes of the ellipse, respectively. The in-plane shear flow $v_{y0}(x')$ develops in and around the electron current sheet.}
\end{figure}

An advantage of using the EMHD model with its simplified assumptions is that only the electron shear flow instability, triggered by electron inertia, can grow in the simulations. Two shear flow structures, in the in-plane and out-of-plane electron flow, developed in the simulations. A simplified schematic representations of the shear flows in the simulations is shown in Fig. \ref{fig:two_shear_flow}. The $x'$-$y'$ axes are along the minor and major axes of the elliptic current sheet, respectively. Although the in-plane flow $v_{y0}(x')$  in the simulations is in the direction making a small angle with the $y'$-axis, it is shown in Fig. \ref{fig:two_shear_flow} to be along the $y'$-axis for simplicity.
In the simulations, the half-thicknesses of the shear layers of the in-plane and out-of-plane electron flows are of the order of an electron inertial length. Ignoring the in-plane shear flows,  the fastest growing mode feeding on the out-of-plane electron shear flow $v_{z0}(x')$ in the presence of a large guide magnetic field is non-tearing with a growth rate $\gamma_f \sim 0.05 \,\omega_{ce}$ and wave numbers $k_yd_e \sim 10$ and $k_zd_e \sim 1$ \cite{jain2015}. However, the simulations do not show any mode developing in the z-direction, i.e., $k_z=0$, suggesting no role of the instabilities feeding on out-of-plane electron shear flow.  Ignoring the out-of-plane electron shear flow, the in-plane flow changes its sign along the $x'$-axis and is thus susceptible to the electron Kelvin-Helmholtz instability (EKHI). We notice that the direction of magnetic field generated by the in-plane shear flow matches that of the external magnetic field $B_z^{ext}$ here. We numerically solved the linear eigenvalue equations of EMHD for a shear flow profile $\mathbf{v}_{y0}(x')=\tanh(x'/L)\hat{y}'$ with a large external magnetic field $B_z^{ext}=18.75$. 
The growth rate is finite only for $k_z=0$ modes, consistent with the simulations in which modes also do not develop in the z-direction. Thus, in-plane shear flow driven instabilities seems to be driving the fluctuations in the simulations.


Our results, obtained in the absence of any ion dynamics, suggest that the electron dynamics play a key role not only in the formation of the electron current sheet but also in the generation of the electromagnetic fluctuations through electron shear flow instabilities in the experiments. Beyond that, for a more precise comparison of the turbulence, one needs to take into account the ion dynamics and its coupling to electrons. 


\begin{acknowledgments}
The work of the authors NJ, AvS and PM was supported by Max-Planck/Princeton Center for Plasma Physics (Project MMC Aero-8003).
\end{acknowledgments}

\bibliography{references}
\end{document}